\documentstyle[12pt]{article}
\font\srm=cmr9

\def\nabl{\nabla\!}
\def\beq{\begin{equation}} \def\eeq{\end{equation}} \def\eqn{\label}
\def\N{^\ast\!n} \def\H{{\cal H}} \def\M{{\cal M}}

\def\en{e^{\rm n}} \def\ep{e^{\rm p}} \def\ee{e^{\rm e}}
\def\eX{e^{_{\rm X}}} 
\def\pin{\pi^{\rm n}} \def\pip{\pi^{\rm p}} \def\pie{\pi^{\rm e}}
\def\piX{\pi^{_{\rm X}}} \def\piU{\pi^{_{\mit \Upsilon}}}
\def\wn{w^{\rm n}} \def\wp{w^{\rm p}} \def\we{w^{\rm e}}
\def\wX{w^{_{\rm X}}} \def\wU{w^{_{\mit \Upsilon}}} 
\def\fn{f^{\rm n}} \def\fp{f^{\rm p}} \def\fe{f^{\rm e}}
\def\fX{f^{_{\rm X}}} \def\fU{f^{_{\mit \Upsilon}}}

\def\phin{\varphi^{\rm n}} \def\phip{\varphi^{\rm p}}
\def\phiU{\varphi^{_{\mit \Upsilon}}}
\def\Pf{{\mit\Phi}} \def\Pfn{{\mit\Phi}_{\rm n}} \def\Pfp{{\mit\Phi}_{\rm p}}
\def\Phin{\Phi_{\rm n}} \def\Phip{\Phi_{\rm p}}
\def\PhiU{\Phi_{_{\mit \Upsilon}}} \def\PhiD{\Phi_{_\Delta}}

\def\lamn{\lambda_{\rm n}} \def\lamp{\lambda_{\rm p}}
\def\lamU{\lambda_{_{\mit \Upsilon}}}

\def\nn{n_{\rm n}} \def\np{n_{\rm p}} \def\ne{n_{\rm e}}
\def\nX{n_{_{\rm X}}} 
\def\nU{n_{_{\mit \Upsilon}}} \def\nV{n_{_{\mit \Phi}}}
\def\rhn{\rho_{\rm n}} \def\rhp{\rho_{\rm p}} 
\def\rhnn{\rho_{\rm nn}} \def\rhnp{\rho_{\rm np}}
\def\rhpn{\rho_{\rm pn}} \def\rhpp{\rho_{\rm pp}}
\def\rhU{\rho_{_{\mit \Upsilon}}} 
\def\rhUV{\rho_{_{\mit \Upsilon \Phi}}}

\def\bn{b_{\rm n}} \def\bp{b_{\rm p}}
\def\bU{b_{_{\mit \Upsilon}}}
\def\xin{\xi_{\rm n}} \def\xip{\xi_{\rm p}} \def\xie{\xi_{\rm e}}
\def\xiX{\xi_{_{\rm X}}} \def\xiU{\xi_{_{\mit  \Upsilon}}}

\def\mun{\mu^{\rm n}} \def\mup{\mu^{\rm p}} \def\mue{\mu^{\rm e}}
\def\muX{\mu^{_{\rm X}}} 
\def\muU{\mu^{_{\mit  \Upsilon}}} \def\muV{\mu^{_{\mit \Phi}}}
\def\nun{\nu^{\rm n}} \def\nup{\nu^{\rm p}} \def\nuU{\nu^{_{\mit \Upsilon}}}
\def\KuUV{{\cal K}^{_{\mit\Upsilon\Phi}}}
\def\Kunn{{\cal K}^{\rm nn}} \def\Kunp{{\cal K}^{\rm np}}
 \def\Kupp{{\cal K}^{\rm pp}}
\def\Kdnn{{\cal K}_{\rm nn}} \def\Kdnp{{\cal K}_{\rm np}}
\def\Kdpn{{\cal K}_{\rm pn}} \def\Kdpp{{\cal K}_{\rm pp}}
\def\KdUV{{\cal K}_{_{\mit\Upsilon\Phi}}}

\def\vsn{\upsilon^{\rm n}} \def\vsp{\upsilon^{\rm p}}
\def\vsU{\upsilon^{_{\mit \Upsilon}}} \def\vsV{\upsilon^{_{\mit \Phi}}} 

\def\alpn{\alpha^{\rm p}_{\rm n}}
\def\omeg{\Omega}

\def\xie{\xi_{\rm e}}

\begin{document}

\title
{Relativistic models for Superconducting-Superfluid Mixtures}

\author { {\bf Brandon Carter and David Langlois}
\\ \hskip 1 cm\\
D\'epartement d'Astrophysique Relativiste et de Cosmologie,\\
Centre National de la Recherche Scientifique, \\Observatoire de
Paris, 92195 Meudon, France.}

\date{\it 7 May 1998}

\maketitle

{\bf Abstract:}
The material below the crust of a neutron star is understood to be describable
in terms of three principal independently moving constituents, identifiable as
neutrons, protons, and electrons, of which the first two are believed to form
mutually coupled bosonic condensates. The large scale comportment of such a
system will be that of a positively charged superconducting superfluid in a
negatively charged ``normal'' fluid background. As a contribution to the
development of the theory of such a system, the present work shows how, subject
to neglect of dissipative effects, it is possible to set up an elegant
category of simplified but fully relativistic three-constituent
superconducting superfluid models whose purpose is to provide realistic
approximations for cases in which a strictly conservative treatment is
sufficient. A ``mesoscopic"  model, describing the fluid 
between the vortices, is constructed, as well as a ``macroscopic" model 
taking into account the average effect of quantised vortices.

\vfill\eject

\section{Introduction}
\label{Sec1}

According to the generally accepted understanding, apart from an outer
magnetosphere of relatively negligible mass and perhaps also, in the most
massive examples, of an inner core consisting of matter in some more or less
exotic state at more than nuclear density, a neutron star can be described in
terms of three principle layers: as well as the degenerate  gas 
of electrons (and muons) that
is present throughout, the outside layer consists of a ``dry'' solid upper 
crust formed just by a lattice of positively charged ions in a degenerate
electron gas; the middle layer consists of a ``wet'' lower crust, whose
solid ionic lattice is interpenetrated by a neutron superfluid; finally the
interior region consists of a neutron superfluid in which the ionic lattice
has dissolved to form a superconducting proton fluid.   The purpose of the
present article is to develop a simplified model that can be used to represent
the dynamics of this superconducting superfluid interior, which constitutes
the greater part of the mass of a typical middlesized neutron star.

For purposes for which only a very crude description of the bulk motion of the
interior superconducting superfluid region is needed, it may be sufficient to
use a model involving only  two independently moving constituents, of which one
represents the neutron superfluid while the other represents everything else,
since short range electromagnetic interactions will tend to ensure that the
positively charged protons and the negatively charged electrons will tend to
move together as an approximately rigid background. One of the advantages of
such a two constituent description is that it is also applicable to the crust
layers, so that -- in an approximation in which details such as the small
anisotropies due to the solidity of the crust are neglected -- it can be used
globally to provide a continuously unified description of the star as a whole.
A suitable two-constituent fluid model for this purpose has recently been
constructed\cite{LSC97} in a relativistic form that includes allowance for
``transfusion'', meaning the slow transfer of baryonic matter (microscopically
attributable to processes such as beta-decay of neutrons to protons) between
the neutron superfluid and the other ``normal'' constituent.

Although adequate for many purposes, the two-constituent fluid
treatment\cite{LSC97} that has just been alluded to is insufficient for the
detailed description of phenomena such as rotation frequency ``glitches'' --
which are observable via the pulsar phenomena -- that depend on the small
deviations from isotropy that occur not only in the crust (as a consequence of
its solidity) but also in the interior, due to the effect of vorticity
quantisation and also (at least in the cases that are observable via the
pulsar phenomena) to the presence of large scale magnetic fields. Using an
variational approach based on the use of a Kalb-Ramond gauge potential in the
manner originally introduced for the treatment of ordinary barotropic perfect
fluids\cite{C94a}, the present authors have previously shown\cite{CL95c} how
to construct a kind of model that -- at least over short and moderate
timescales for which ``transfusion'' processes can be neglected -- can provide
a satisfactorily relativistic macroscopic description of the effect of the
anisotropy due to fibration by quantised vortices in what is describable, at a
``mesoscopic'' (submacroscopic but more than microscopic) scale outside the 
vortices, by a simple superfluid model. However although this kind of
anisotropic superfluidity model is directly applicable to the terrestrially
familiar example of liquid helium, the utility for neutron stars of such a
single-constituent representation is limited by its neglect of the
electromagnetically interacting constituents, namely the electrons and the
protons, which together represent only a small fraction of the mass density,
but which nevertheless play an essential role in phenomena involving magnetic
effects. 

It has already been shown that the variational treatment in terms of 
Kalb-Ramond potentials can be straightforwardly generalised from the original
single-constituent perfect fluid case\cite{C94a} to the case of perfect
two-constituent electrically conducting fluid models\cite{C94b} (including the
special kind -- with the negatively charged constituent restricted to be
irrotational -- that can be used for a ``mesoscopic'' description of
the experimentally familiar kind of type II electronic superconductor using a
crude but fully relativistic representation that neglects the anisotropy due
to the solid, not fluid, structure of the positively charged ionic background
constituent). Following this example, the purpose of the present work  is to
show how  -- still neglecting the ``transfusion'' processes that are likely to
be important only in the very long run -- our previous macroscopic
model\cite{CL95c}, representing the effect of vortex fibration in a simple
superfluid, can be extended to a three-constituent model providing a
description that is applicable to a superconducting superfluid in which the
other two independently moving constituents are the superfluid protons and the
degenerate electron gas.

A considerable amount of work on three-constituent models for this purpose has
already been carried out by several authors\cite{VS81,MenLin91,Men91,SS95}, but
until now only within the framework of a ``non-relativistic'' Newtonian
description (as a generalisation of the kind of two-constituent model used for
treatment of an ordinary ``type II'' metallic superconductors of the sort
familiar in a laboratory context, for which the two constituents are
superconducting electrons and a solid ionic background lattice).  Some of this
previous work includes allowance for details such as dissipative effects e.g.
to various kinds of resistivity, that in the present description will for the
sake of clarity will be provisionally left aside for treatment in a subsequent
treatment. The omission of such dissipative effects is however by no means the
only reason why the conservative treatment given here turns out to be
technically simpler and more elegant than its ``non-relativistic''
predecessors: the other reason is that the Galilei invariance group applicable
to the Newtonian description is more complicated than (since it is a
degenerate limit of) the (in the technical sense ``semi-simple'') Lorentz
invariance group that is locally applicable to the fully (general as well as
special) relativistic treatment provided here. When the electromagnetic 
coupling is taken into account, the relativistic description is even more 
satisfying in that the material part and the electromagnetic part 
are treated on the same footing instead of the awkward mixture of 
Galilean and Lorentzian invariances in the Newtonian context.

It is to be remarked that the improvement not just in physical accuracy but
also in mathematical elegance that is obtainable by going from a Newtonian to
a relativistic treatment has already been demonstrated in the case of Landau's
original two-constituent superfluid model (in which the second constituent just
represents the entropy current), whose relativistic generalisation\cite{CK92b}
has provided insights that have lead to a more efficient ``canonical''
description\cite{CK94} of its original Newtonian version. This suggests, as a
challenge for future work, that the traditional presentation of the more
elaborate three-constituent fluid models developed in previous
work\cite{MenLin91,Men91,SS95}, should also be susceptible to similar
improvements in elegance by an analogous ``canonical'' reformulation along
lines that would presumably be obtainable by taking an appropriate Newtonian
limit of the even more elegant relativistic kind of model developed here.

Of greater importance than the analysis of the Newtonian limit, as a challenge
for future work, the kind of model developed here  will need to be completed
by  the provision (on the basis of appropriate microphysical analysis) of
explicit forms and values for the functions and parameters involved , while it
will also be necessary to develop the more extended treatment needed to
include suitable allowance for dissipative effects (not to mention refinements
such as allowance for the spinor as opposed to scalar nature of the Cooper
pairs involved in the microscopic description\cite{PinesAlpar85} of both the
neutron superfluid and the superconducting proton current).  

Another challenge is the development of a corresponding relativistic
description for the solid crust layers. An appropriate relativistic formalism
for describing the purely solid ``dry'' upper crust layer of a Newton star
was developed quite a long time ago\cite{CQ72,C73,CQ75}, though its technical
complexity has delayed its effective application to specific astrophysical
problems\cite{Priou92}. What has yet to be constructed is the even more
elaborate generalisation required for treating the ``wet'' lower layer of the
solid crust: machinery capable of providing a treatment on a ``mesoscopic''
scale of the interaction between the permeating neutron superfluid and the solid
background lattice has already been made available\cite{C89}, but it still
remains to develop the corresponding macroscopic analysis required for the
relativistic description of the large scale effects of phenomena such as
vortex pinning\cite{AAPS84} in such a ``wet'' solid.

As declared above, the ultimate task of the present work is to provide a kind
of model that can account for the macroscopic anisotropies attributable to
quantised vortex lines associated with both rotation and magnetic fields, in
the deep superconducting superfluid interior of a neutron star. However, as a
preliminary step, we must first develop a suitably relativistic ``mesoscopic''
(submacroscopic but not microscopic) scale treatment applicable to the
superconducting superfluid in between the relevant stringlike vortex defects.
The most elegant versions of the models presented below are those
that apply to the zero temperature limit, which should be a good enough
approximation for most relevant astrophysical applications in neutron stars.
We shall however include a rudimentary allowance for the effect of a finite
temperature associated with a non-vanishing conserved
entropy distribution comoving with the degenerate electrons
in the negatively charged ``normal'' constituent.

\section {Three constituent perfectly conducting fluids.}
\label{Sec2}

Following what has long been established as a routine procedure\cite{C85} --
(the natural relativistic analogue of the phenomenological approach
developed in a Newtonian context \cite{AndreevBashkin76} as a generalisation
of Landau's original two-constituent model) -- the task of the present section
is to set up the category of simple three-constituent perfectly conducting
 fluid models that is needed as a preliminary basis for the more
specific developments that follow. As the specialisation to be treated in the
next section, this category includes the particular restricted case that is
appropriate for the treatment of the kind of superconducting superfluid that
is relevant in neutron stars on a ``mesoscopic'' scale, meaning a scale large
compared with that of the underlying microscopic particle description, but
small compared with the macroscopic scale of separation between the vortex
defects on which the superfluid comportment is locally violated. On the other
hand the category set up in the present section is itself a just specially
simple limit within the more general category to be developed in the later
sections for the purpose of treating the superconducting superfluid on a
``macroscopic'' scale, meaning a scale that is large compared with the
separation between vortices.

The three independent constituents under consideration are identifiable as the
superfluid neutrons -- which make up most of the mass density in the relevant
neutron star layers -- with number current four-vector $\nn^{\,\rho}$, the
superconducting protons, -- which make up a small but significant part of the
mass density -- with number current four-vector $\np^{\,\rho}$, and finally
the degenerate non-superconducting background of electrons with number current
vector $\ne^{\,\rho}$.  The latter make up a negligibly small fraction of the
mass density, but they nevertheless have a crucially important role in so far
as electomagnetic effects are concerned, since in terms of the electron charge
coupling constant $e$ the corresponding total electric current vector four
vector will be given by 
 \beq 
j^\mu=e\big(\np^{\,\rho}-\ne^{\,\rho}\big) \, .
 \eqn{1}\eeq

As well as the three principal constituents that have just been listed, our
discussion in this section will also include allowance for a fourth
constituent, namely the entropy. However this will not be treated as fully
independent, since (as will be a realistic approximation for most of the
relevant astrophysical applications) its current four-vector $s^\mu$ will be
postulated to be constrained so as to be aligned with that of the electrons.
This means that, apart from the ``super'' constituents with currents
$\nn^{\,\rho}$ and $\np^{\,\rho}$, there will be just a single independently
moving ``normal'' constituent with current in the direction specified by a
unit vector $u^\rho$, as characterised by the normalisation condition
 \beq 
u^\rho u_\rho=-1\, ,
 \eqn{2}\eeq
(on the usual understanding that the units are such that the speed of light
is unity) with
 \beq
\ne^{\,\rho}=\ne u^\rho\, ,              \eqn{3}\eeq
 \beq
s^\rho = s u^\rho\, , 
 \eqn{4}\eeq
where the scalars $s$ and $\ne$ are respectively the entropy density and
the electron number density in the ``normal'' rest frame specified by
$u^\rho$.

Since our present treatment will be restricted to the conservative limit in
which dissipative effects are neglected, the analysis will be assumed to be
expressible in terms of a variational principle based on a Lagrangian density
that, in the absence of electromagnetic effects, can be presumed to be given
by a function $\Lambda_{\rm M}$ say depending just on the six independent
scalars obtainable by mutual contractions of the three independent current
vectors $\nn^{\,\rho}$, $\np^{\,\rho}$, $\ne^{\,\rho}$ and on the entropy
density $s$. The independent variations of these quantities determine the
effective momentum covectors $\mun_\mu$ and $\mup_\mu$ respectively associated
with the superfluid neutrons and the superconducting protons, together with
the ``normal'' momentum covector $\mue_{\,\mu}$ associated with the electrons,
and the corresponding temperature $\Theta$, via a prescription of the form
 \beq
\delta\Lambda_{\rm M}=\mun_{\ \rho}\, \delta\nn^{\,\rho} +\mup_{\ \rho}\, 
\delta\np^{\,\rho}+ \mue_{\ \rho}\, \delta\ne^{\,\rho}-\Theta\, \delta s 
+{\partial \Lambda_{\rm M} \over \partial g_{\rho\sigma}}\,
\delta g_{\rho\sigma} \, ,
 \eqn{5}\eeq
in which, as an elementary Noether type identity, we shall automatically
have
 \beq
{\partial \Lambda_{\rm M} \over \partial g_{\rho\sigma}}=
{\partial \Lambda_{\rm M} \over \partial g_{\sigma\rho}}= 
{_1\over ^2}\big( \mun_{\ \nu}\, \nn^{\,\rho} +\mup_{\ \nu}\, 
\np^{\,\rho}+ \mue_{\ \nu}\, \ne^{\,\rho}\big)\, g^{\nu\sigma}\, .
 \eqn{6}\eeq
It is useful to allow here for the possibility of varying the background
spacetime metric $g_{\rho\sigma}$, not only for the purpose of dealing with
cases in which one may be concerned with General Relativistic gravitational
coupling, but even for dealing with cases in which one is concerned only with
a flat Minkowski background, since, as will be made explicit below, the effect
of virtual variations with respect to the relevant curved or flat background
metric can be used for evaluating the relevant ``geometric'' stress energy 
momentum density tensor $T^{\rho\sigma}$. 

The preceding formulae (\ref{5}) and (\ref{6}) can conveniently be combined
in a single expression of the more concise form 
 \beq
\delta\Lambda_{\rm M}= \muX_{\ \rho}\, \delta\nX^{\,\rho}-\Theta\, \delta s
+{_1\over^2} \muX_{\ \nu}\, \nX^{\,\rho}\,
g^{\nu\sigma}\, \delta g_{\rho\sigma}\, ,
\eqn{7}\eeq
using the summation convention for ``chemical'' indices represented
by capital Latin letters running over the three relevant values, namely
{\srm X}=n,p,e. Using this convention, the 
equation (\ref{1}) for the electric current density can be rewritten
in the concise form
 \beq
j^\rho=\eX \nX^{\,\rho} \, ,
 \eqn{8}\eeq
where the charges per neutron, proton, and electron are given 
respectively by  $\en=0$, $\ep=e$, and $\ee=-e$.

The standard minimal prescription for inclusion of electromagnetic interactions
is to use a combined Lagrangian scalar density in which the ``matter''
contribution $\Lambda_{\rm M}$ is augmented by an electromagnetic field
contribution $\Lambda_{\rm F}$ and a gauge dependent coupling term of 
the usual form to give a total Lagrangian scalar ${\cal L}$ expressible 
as
 \beq
{\cal L}=\Lambda + j^\rho A_\rho \, ,
 \eqn{9}\eeq
where the gauge independent part has the form
 \beq
\Lambda=\Lambda_{\rm M}+\Lambda_{\rm F}\, ,
 \eqn{10}\eeq
and where $A_\rho$ is the electromagnetic gauge form, while the gauge 
independent contribution $\Lambda_{\rm F}$ is given in terms of the 
corresponding electromagnetic field tensor,
 \beq
F_{\rho\sigma}=2\nabl_{[\rho}A_{\sigma]}\, ,
 \eqn{11}\eeq
(using square brackets to indicate index antisymmetrisation) by
the standard Maxwellian formula
 \beq
\Lambda_{\rm F}={1\over 16\pi} F_{\rho\sigma}F^{\sigma\rho}\, .
 \eqn{12}\eeq
The variation of the total will then be given by
 \beq
\delta{\cal L}=\piX_{\ \rho}\,\delta \nX^{\,\rho}-\Theta\,\delta s
+j^\rho\,\delta A_\rho+{1\over 8\pi}F^{\sigma\rho}\,\delta F_{\rho\sigma} 
+{1\over 2}\Big(\muX_{\ \nu}\, \nX^{\,\rho}\, g^{\nu\sigma} 
+{1\over 4\pi}F^{\nu\rho}F_\nu^{\ \sigma}\Big)\,\delta g_{\rho\sigma}\, ,
 \eqn{13}\eeq
where the gauge dependent total momentum covectors are given by
 \beq
\piX_{\ \rho}=\muX_{\ \rho}+\eX A_\rho\, .
 \eqn{14}\eeq

In order to characterise a variation principle that can be used to specify the
equations of motion of a multiconstituent medium, the electromagnetic
potential covector $A_\rho$ is of course allowed to vary freely, but the
variations of the independent currents $\nX^{\rho}$ cannot be allowed to be
arbitrary, since this would evidently lead to an overdetermined system in
which the momenta would simply have to vanish. The standard procedure for
restricting the allowed current variations so as to obtain a system with the
appropriate number of degrees of freedom for an ordinary fluid mixture is to
require that the the current variations be determined by free displacements of
the corresponding worldlines -- in the manner originally introduced for the
particular case of a relativistic perfect fluid by Taub\cite{Taub54}. The
ensuing variational equations then automatically ensure the conservation
under transport by each current flow $\nX^{\,\rho}$ of
the corresponding generalised vorticity tensor as defined by
 \beq
\wX_{\,\rho\sigma}=2\nabl_{[\rho}\piX{_{\!\sigma]}}
=2\nabl_{[\rho}\muX{_{\!\sigma]}}+\eX F_{\rho\sigma}\, .
 \eqn{15} \eeq
Such an exterior derivative will of course automatically be ``closed'',
i.e. its own exterior derivative will vanish:
\beq
\nabl_{[\nu}\,\wX\!{_{\rho\sigma]}}=0\, .
 \eqn{16} \eeq 

Within such a perfect multiconstituent fluid category, the special properties
of superfluidity and superconductivity are characterised by the constraint
that the corresponding conserved vorticity should vanish initially, and hence,
since it is conserved, throughout the  ensuing evolution of the system.
Such constraints -- which in our present application concern the neutrons and
the protons but not the electrons -- are interpretable as integrability
conditions for corresponding potential scalars. For the neutrons we shall
have
 \beq
\wn_{\,\rho\sigma}=0\,  \Rightarrow\, \pin_{\ \rho} \equiv \mun_{\ \rho}
={\hbar\over 2} \nabl_\rho\,\phin\, ,
 \eqn{19}\eeq
while for the protons we shall have
 \beq
\wp_{\,\rho\sigma}=0\, \Rightarrow \, \pip_{\ \rho}
\equiv  \mup_{\ \rho}+eA_\rho={\hbar\over 2} \nabl_\rho\,\phip\, ,
 \eqn{20}\eeq
where the locally defined potentials $\phin$ and $\phip$ are interpretable as
phase angles (with period $2\pi$) associated with underlying boson
condensates, and the factors 2 in the denominators are included to allow for
the fact that the relevant bosons are presumed to consist not of single
protons or neutrons but of Cooper type pairs.

The way the Taub type ``convective'' variational procedure works out is
as follows. For any current vector $n^{\rho}$ the corresponding variation 
rule is interpretable as meaning that the ensuing variation of
its dual three-form, as given -- in terms of the antisymmetric measure tensor
$\varepsilon_{\mu\nu\rho\sigma}$ associated with the 4-dimensional spacetime
metric $g_{\mu\nu}$ -- by
  \beq
\N_{\mu\nu\rho}=\varepsilon_{\mu\nu\rho\sigma}n^{\sigma}\, ,
 \eqn{21}\eeq
should be specified by its Lie derivative 
with respect to the relevant worldline displacement vector field,
$\xi^{\rho}$ say, i.e. 
 \beq
\delta  \N_{\mu\nu\rho} =\xi^{\sigma}\nabl_\sigma\, \N_{\mu\nu\rho}
+ 3\N_{\sigma[\mu\nu}\nabl_{\rho]}\xi^{\,\sigma}\, ,
 \eqn{22}\eeq
where $\nabl_\mu$ denotes the covariant differentiation operator specified
by the relevant (flat or curved) spacetime background metric $g_{\mu\nu}$.

When applied to the particular case of the ``normal'' (i.e.
non-superfluid) electron current $\ne^{\,\rho}$, for a corresponding
displacement vector field $\xie^{\,\rho}$, the foregoing prescription
can be seen to give a variation of the form 
 \beq 
\delta \ne^{\rho}=\xie^{\,\sigma}\nabl_\sigma\,\ne^{\,\rho}-\ne^{\,\sigma}
\nabl_\sigma\,\xie^{\,\rho} +\ne^{\,\rho}\nabl_\sigma\,\xie^{\,\sigma}\, 
-{_1\over^2}\,\ne^{\,\rho}\, g^{\mu\nu}\,\delta g_{\mu\nu}\, , 
\eqn{23}\eeq
from which it follows that the variation of the scalar electron number density
introduced in (\ref{3}) will be given by
 \beq
\delta \ne= \xie^\sigma\nabl_\sigma\, \ne+
\ne u_\rho u^\sigma \nabl_\sigma\,\xie^\rho+\ne \nabl_\sigma\,\xie^{\,\sigma}
-{_1\over^2}\ne\big(g^{\mu\nu}+u^\mu u^\nu\big)\,\delta g_{\mu\nu}\, ,
 \eqn{24}\eeq
Since we are postulating that the electron current is comoving with the
entropy current as part of a combined ``normal'' constituent, for which the
shared unit 4-vector appearing in (\ref{3}) and (\ref{4}) will thus be subject
to the variation law
 \beq
\delta u^\rho=\xie^{\,\sigma}\nabl_\sigma\, u^\rho 
- u^\sigma\nabl_\sigma\,\xie^{\,\rho} 
- u^\rho u_\nu u^\sigma\nabl_\sigma\,\xie^{\,\nu}
+{_1\over^2} u^\rho u^\mu u^\nu\,\delta g_{\mu\nu}\, ,
 \eqn{25}\eeq
it follows that the variation of the entropy current will be governed by the
same displacement vector field $\xie^{\,\rho}$ as that of the electron
current, and hence the variation of the entropy density introduced in
(\ref{4}) will be governed by a rule of the same form as (\ref{24}), namely
 \beq
\delta s= \xie^{\,\sigma} \nabl_\sigma\, s+ s u_\rho u^\sigma 
\nabl_\sigma\,\xie^{\,\rho} +s \nabl_\sigma\, \xie^{\,\sigma}
-{_1\over^2} s \big(g^{\mu\nu}+u^\mu u^\nu\big)\,\delta g_{\mu\nu}\, .
 \eqn{26}\eeq

When metric variations are involved, the application of the variation
principle requires allowance not just for the variation of the Lagrangian
scalar ${\cal L}$, but also for the variation of the relevant spacetime
measure, which is proportional to the square root of the modulus of the metric
determinant $\vert g \vert$. This means that the complete variational
integrand that one needs to evaluate is the ``diamond'' variation as defined
by
 \beq
\diamondsuit {\cal L}=\Vert g \Vert^{-1/2}\, \delta\Big(\Vert g \Vert^{1/2}
{\cal L} \Big) = \delta{\cal L} +{_1\over^2} {\cal L} g^{\mu\nu}\, 
\delta g_{\mu\nu}\, .
 \eqn{27}\eeq

By taking the variations of the neutron and proton currents to be given in the
same way as that of the electron current by corresponding displacement vector
fields $\xin^{\,\rho}$ and $\xip^{\,\rho}$, the required ``diamond'' variation
of the Lagrangian scalar density given by (\ref{9}) can be obtained in the
standard form
 \beq
\diamondsuit {\cal L} =\fX_{\,\rho}\, \xiX^{\,\rho}+\Big(j^\rho-
{1\over 4\pi}\nabl_\sigma\, F^{\rho\sigma}\Big)\,\delta A_\rho
+{_1\over^2} T^{\mu\nu}\,\delta g_{\mu\nu} 
+\nabl_\sigma\, {\cal R}^\sigma \, ,
 \eqn{28}\eeq
in which the coefficients $\fX_{\,\rho}$ of the independent displacement
displacement vector fields $\xiX^{\,\rho}$ are respectively
interpretable as the force densities acting on the corresponding
constituents, while the (symmetric) tensor $T^{\mu\nu}$ evidently represents 
the ordinary (``geometric'') stress momentum energy density. 
It can be seen that the force densities acting on the superfluid neutrons 
and the superconducting protons will be given respectively by 
 \beq
\fn_{\,\rho}= \nn^{\,\sigma} \wn_{\,\sigma\rho}
+\mun_{\,\rho} \nabl_\sigma\,\nn^{\,\sigma}\, ,
 \eqn{29}\eeq
and
 \beq
\fp_{\,\rho}= \np^{\sigma}\wp_{\,\sigma\rho}
+\pip_{\,\rho} \nabl_\sigma\,\np^{\,\sigma}\, .
 \eqn{30}\eeq
where $\wn_{\,\sigma\rho}$ and $\wp_{\,\sigma\rho}$ are the corresponding
vorticity two-forms as defined by (\ref{15}). The force 
density acting on the ``normal'' constituent -- consisting of the electrons
and the comoving entropy current -- can be seen to be given by an expression
of the not quite so simple form
 \beq
\fe_{\,\rho}= 2u^\sigma \nabl_{[\sigma}\,\Pi_{\rho]}+\Pi_\rho\nabl_\sigma\,
 u^\sigma + u^\sigma\pie_\sigma \nabl_\rho\, \ne-\Theta\nabl_\rho\,s \, ,
 \eqn{31}\eeq
in which the one-form $\Pi_\rho$ is the effective momentum density
of the normal constituent (electrons and entropy) treated as a whole,
as given by
 \beq
\Pi_\rho=\ne\pie_{\,\rho}+s\Theta u_\rho\, .
 \eqn{32}\eeq
The stress momentum energy density tensor can be seen to have the form
\beq
T^{\rho\sigma}=T_{\rm M}^{\ \rho\sigma}+T_{\rm F}^{\ \rho\sigma}\, ,
\eqn{33}\eeq
where the part derived just from the material Lagrangian density 
contribution $\Lambda_{\rm M}$ is given by
\beq
T_{{\rm M}\ \, \sigma}^{\ \rho}=\nX^{\,\rho}\muX_{\,\sigma}+
s\Theta u^\rho u_\sigma +\Psi_{\rm M} g^\rho_{\,\sigma}\, ,\hskip 1 cm
\Psi_{\rm M}=\Lambda_{\rm M}-\nX^{\,\sigma}\muX_{\,\sigma}+s\Theta\, ,
\eqn{34}\eeq
while the electromagnetic contribution has the usual Maxwellian form 
 \beq
T_{{\rm F}\ \, \sigma}^{\ \rho}={1\over 4\pi}\Big( F^{\nu\rho} F_{\nu\sigma}
-{1\over 4}F^{\mu\nu}F_{\mu\nu} g^\rho_{\,\sigma}\Big)\, .
 \eqn{35}\eeq
Although it is irrelevant as far as the variation principle is concerned,
it is to be noted for the sake of completeness that the remainder
current in the final divergence term of (\ref{28}) will be given by
 \beq
{\cal R}^\sigma=2\piX_{\, \nu}\xiX^{\,[\sigma}\nX^{\,\nu]}+2\Theta u_\nu 
\xie^{\,[\sigma} s^{\nu]} +{1\over 4\pi}F^{\nu\sigma}\,\delta A_\nu \, .       
 \eqn{36}\eeq

It is evident from (\ref{28}) that the requirement of invariance of the
action with respect to free variations of the gauge
form $A_\rho$ leads to electromagnetic source equations of the usual
Maxwellian form
 \beq
\nabl_\sigma\, F^{\rho\sigma}=4\pi j^\rho\, .
 \eqn{37}\eeq
Subject to this condition, it follows as a Noether type 
identity (obtainable by identifying all the displacement vectors with
a single vector field with respect to which the system is displaced as a 
whole\cite{C89,C91}) that, independently of the other
field equations one must have a force balance relation of the form
 \beq
\fn_{\,\rho}+\fp_{\,\rho}+\fe_{\,\rho}=\nabl_\sigma\, T^\sigma_{\ \rho}\, ,
 \eqn{38}\eeq
in  which the right hand side represents the total force density acting on the
medium, which will of course vanish for the kind of isolated strictly
conservative model with which we are concerned here. More particularly, it is
evident from (\ref{28}) that the content of the  Taub type ``convective''
variation principle expressed by the condition that independent adjustments of
the displacement fields should have no effect is equivalent to the dynamical
requirement that each of the separate force density contributions on the left
of (\ref{38}) should vanish separately: in short the equations of motion of
the system will be expressible simply by the condition
\beq
 \fX_{\,\rho}=0
\eqn{39}\eeq 
for each of the three relevant chemical index values {\srm X}=n,p,e. 

Assuming that (unlike what would be required for a model of
``transfusive'' type\cite{LSC97}) there is no algebraical constraint on the
momenta, it can be seen (by contracting the relevant force formulae
(\ref{29}) and (\ref{30}) with the corresponding current vectors)
that for the cases of the neutrons and
the protons the conditions (\ref{39}) include the implication that their
currents are separately conserved,
 \beq
\nabl_\rho \,\nn^{\,\rho}=0 \, ,\hskip 1 cm
\nabl_\rho\, \np^{\,\rho}=0 \, .
 \eqn{40}\eeq
The remaining contents of the neutron and proton equations of motion  will
simply reduce to the respective forms
 \beq \nn^{\,\sigma}\wn_{\,\sigma\rho}=0 \, , \hskip 1 cm
\np^{\,\sigma}\wp_{\,\sigma\rho}=0 \, ,
 \eqn{41}\eeq
where the vorticity two forms are as defined by (\ref{15}). It is well
known that the form of such equations ensures that, in view of the
closure property (\ref{16}), the neutron and proton vorticity forms
will be conserved in the sense of being dragged along by the corresponding
flows\cite{C89}, which means that they are consistent with the
special requirements for superfluidity and superconductivity respectively,
which are that they should simply vanish, in accordance with (\ref{19})
and (\ref{20}). 

For the non-superfluid ``normal'' constituent, the implications of the zero
force condition (\ref{39}) are not so simple. In this case, it can be seen
from (\ref{31}) that contraction with the relevant unit flow tangent vector 
will just give
 \beq
u^\rho\pie_{\,\rho}\nabl_\sigma\,\ne^\sigma= 
\Theta \nabl_\sigma\, s^\sigma \, .
 \eqn{42}\eeq
To obtain separate current conservation laws that are required, it should be
understood that the electron current is not specified freely but is given by a
procedure of the kind appropriate for non-barotropic perfect fluid
theory\cite{ComerLanglois94}  via a projection (of the kind that is standard as the
basis of relativistic elasticity theory\cite{CQ72,C73}) onto a three
dimensional base manifold endowed with its own measure. This means that in
terms of the space-time pull-backs of suitably chosen base coordinates
$X^{_A}$, $A$=1,2,3, it will be given, using the abbreviation 
$X^{_{\!A}}_{\,\mu}=\nabl_\mu X^{_{\!A}}$, 
by an expression of the form $\ne^{\,\mu}=$
$\varepsilon^{\mu\nu\rho\sigma}X^{_1}_{\ ,\nu} X^{_2}_{\ ,\rho}X^{_3}_{\
,\sigma}$. The variation law (\ref{23}) is equivalent to what is obtained by
allowing free variations of the $X^{_{\!A}}$ considered as a set of three
independent scalar fields on spacetime: such variations are equivalent to the
specification of a corresponding displacement vector field $\xie^{\,\rho}$ for
which one will have $\delta X^{_{\!A}}=\xie^{\,\rho}\nabl_\rho\, X^{\,\rho}$. 
Such a construction automatically ensures that electron current
will be identically conserved, and the dynamical equation (\ref{42})
then entails a similar condition for the entropy, i.e. one ends up
with the separate conservation laws
 \beq
\nabl_\sigma\, \ne^{\,\sigma}=0\, ,\hskip 1 cm
\nabl_\sigma\, s^{\,\sigma}=0\, .
 \eqn{43}\eeq
Under these conditions, the remaining contents of the equations of motion of
the ``normal'' constituent, as expressed by the vanishing according to
(\ref{39}) of the force density $\fe_{\,\rho}$ given by (\ref{31}), will take
the form
 \beq
\ne^{\,\sigma}\we_{\,\sigma\rho} +s^\sigma
\nabl_\sigma\,\big(\Theta u_\rho\big) +s\nabl_\rho\,\Theta=0\, ,
 \eqn{44}\eeq
which -- unlike the corresponding dynamical equations (\ref{41}) for the
neutrons and the protons -- does not have the form of a generalised  vorticity
conservation law, except in the zero temperature limit, $\Theta=0$, for which
only the first term remains.

\section{Mesoscopic superconducting superfluid model.}
\label{Sec3}

Within the category of perfectly conducting fluid motions, as characterised by
the equations of motion (\ref{40}), (\ref{41}), (\ref{43}), (\ref{44}), the
``mesoscopic'' superconducting superfluid case relevant on an intermediate
scale (in between the vortices) will be characterised by the constraints
(\ref{19}) and (\ref{20}), whose compatibility with the dynamical equations is
ensured by the particular form of (\ref{40}). Instead of obtaining them as a
particular case within the general perfectly conducting fluid category, we can
proceed directly to the specialised superconducting superfluid equations of
motion, as given just by (\ref{40}), (\ref{43}), (\ref{44}) together -- in
accordance with (\ref{19}) and (\ref{20}) -- with
 \beq
 2\pin_{\,\rho}={\hbar}\nabl_\rho\,\phin\, ,\hskip 1 cm
2\pip_{\,\rho}={\hbar}\nabl_\rho\,\phip\, ,
 \eqn{47}\eeq
by using a variational procedure of a modified kind. Since this new 
kind of procedure will treat the ``super'' currents on a rather different
footing from the others, its presentation will be facilitated by the
introduction of a correspondingly restricted chemical index, for which we
shall use the capital Greek letters ${\mit\Upsilon}$, ${\mit\Phi}$ which
differ from the unrestricted chemical indices ${\srm X}$, ${\srm Y}$ in having
a range that is restricted to exclude the ``normal'' current values. For the
three-constituent category under consideration here, the only ``normal'' value
to be excluded is ${\srm} X$=e. It follows that, for this category, the Greek
letters will be interpretable as {\it isotopic} indices, taking the pair of
values ${\mit \Upsilon}$=n for the neutron case and ${\mit \Upsilon}$=p
for the proton case. With this convention, the two separate equations
(\ref{47}) can be combined into  the more concise form
 \beq
2\piU_{\,\rho}={\hbar}\nabl_\rho\,\phiU\, ,
 \eqn{48}\eeq
in which it is to be recalled that the factor 2 is present in order that the
left hand side should represent the effective momentum covector of the
relevant bosonic unit, which will consist not of a single baryon but of a
Cooper type pair. 

According to the rules of the modified procedure, instead of being restricted
to satisfy displacement variation rules of the kind (\ref{23}) satisfied by
the ``normal'' constituent, it is permissible for the isotopic doublet formed
by the  ``super'' currents $\nn^{\,\rho}$ and $\np^{\,\rho}$ to vary freely
apart from the restriction that the corresponding conservation laws
(\ref{40}) should be satisfied. The enforcement of these conservation laws,
which can be combined in the single expression
 \beq
\nabl_\sigma\,\nU^{\,\sigma}=0\, ,
 \eqn{49}\eeq
is to be obtained in the usual way by the introduction of corresponding
Lagrange multipliers. Provided they are introduced with the appropriate
normalisation, these multipliers turn out to be identifiable with the
potentials $\phin$ and $\phip$ introduced in (\ref{19}) and (\ref{20}). This
is done my making the replacement
 \beq
{\cal L}\, \mapsto {\cal L}_{\rm I}\, ,
 \eqn{50}\eeq
where the modified Lagrangian scalar is given by the expression
\beq
{\cal L}_{\rm I}={\cal L} +{_{\hbar}\over ^2}\phiU\nabl_\sigma\nU^{\,\sigma}\, ,
 \eqn{51}\eeq
in which summation over the isotopic index ${\mit \Upsilon}$ is of course 
to be understood. The correspondingly modified replacement of the 
``diamond'' variation formula (\ref{28}) is
\vfill\eject 
$$
\diamondsuit {\cal L}_{\rm I} = \big(\piU_{\,\rho}-{_{\hbar}\over^2}
\nabl_\rho\,\phiU \big)\delta\nU^{\,\rho} +\fe_{\,\rho}\, \xie^{\,\rho}
+\Big(j^\rho-{1\over 4\pi}\nabl_\sigma\, F^{\rho\sigma}\Big)\,\delta A_\rho 
 $$\beq
+{_{\hbar}\over^2}\big(\nabl_\sigma\,\nU^{\,\sigma}\big)\delta\phiU 
+{_1\over^2} T^{\mu\nu}\,\delta g_{\mu\nu} 
+\nabl_\sigma\, {\cal R}_{\rm I}^{\ \sigma} \, ,
 \eqn{52}\eeq
in which the force density $\fe_{\rho}$ acting on the normal constituent is
given by exactly the same formula (\ref{31}) as before. Although it is
irrelevant for the purposes of application of the action principle, it may be
noted that instead of (\ref{36}), the remainder current in the final
divergence term of (\ref{52}) will be given by the rather different formula
  \beq 
{\cal R}_{\rm I}^{\ \sigma}={_{\hbar}\over^2}\phiU\big(\delta \nU^{\,\sigma}
+{_1\over^2} \nU^{\,\sigma} g^{\mu\nu}\delta g_{\mu\nu}\big) 
+2\Pi_\nu\xie^{\,[\sigma}u^{\,\nu]}
+{1\over 4\pi}F^{\nu\sigma}\,\delta A_\nu \, .       
 \eqn{54}\eeq
More to the point for practical purposes, the stress energy momentum density
tensor $T^{\mu\nu}$ in (\ref{52}) will be given by the same formulae
(\ref{33}), (\ref{34}), (\ref{35}) as before, apart from the fact that in the
new version the relevant generalised pressure function will be given by
 \beq
\Psi_{\rm M}=\Lambda_{\rm M}-{_{\hbar}\over^2}\nn^{\,\sigma}
\nabl_{\,\sigma}\,\phin + \np^{\,\sigma}\big(eA_\sigma -{_{\hbar}\over^2}
\nabl_{\,\sigma}\,\phip\big) -\ne^{\,\sigma}\mue_{\,\sigma}+s\Theta\, ,
\eqn{55}\eeq
which can be seen to be equivalent to the original version, as given by
(\ref{34}), when the superconducting superfluid field equations (\ref{47}) are
satisfied, as is necessary in this modified formulation for invariance of the
action with respect to arbitrary infinitesimal variations of $\nn^{\,\rho}$
and $\np^{\,\rho}$.

The essential difference between a ``normal'' and a ``super'' current is not
just that the latter is irrotational so that its momentum covector is locally
proportional to the gradient of a scalar field, but more particularly that
this scalar field will be globally definable as a periodic phase angle, so
that the corresponding circulation integral will be quantised.  Specifically,
according to (\ref{48}), the relevant neutron and proton circulation integrals
for a given circuit will be given in the present case by
 \beq
\oint\piU_{\,\rho}\, dx^{\,\rho}=\pi\hbar\,\nuU\, ,
 \eqn{56}\eeq
for a pair of integral proton and neutron phase winding numbers respectively
given by $\nun$ and $\nup$. These numbers will remain constant as the circuit
is continuously displaced, except when it crosses what, on a mesoscopic
scale will be describable as a string like singularity, but what on a
microscopic scale will be a vortex type defect whose core consists of a region
in which the superfluidity property must break down. It is expected that,
in most of the relevant neutron star layers, conditions will be such
as to ensure that such vortex defects will generally be of elementary
type, meaning that they will either be simple neutron flow vortices,
as characterised, for an appropriate choice of orientation,
by winding numbers $\nun= 1$, $\nup=0$, or simple proton vortices, as
similarly characterised by $\nun=0$, $\nup=1$. In order to obtain a 
description that is applicable on a macroscopic scale, it will evidently be
necessary to average over a large number of such (elementary, or in more
exotic circumstances, higher order) vortices. Following the example of our
treatment of the single constituent case\cite{CL95c}, the purpose of the next
two sections is to develop and show how to apply the kind of formalism needed
to achieve this.

\section{Macroscopic allowance for vorticity fibration}
\label{Sec4}

The standard (Taub type) flow line variational procedure illustrated by the
category of models set up in  Section \ref{Sec2} has the advantage of being
extremely versatile: for example it can be generalised for application to
simple elastic solids\cite{C73}, as well as to conducting solids\cite{C89},
and it has been used most recently for setting up the ``transfusive''
two-constituent model\cite{LSC97} that has been designed to provide a crude
but globally applicable description of a neutron star as a whole. However for
the purpose of dealing with the macroscopic effects of vorticity quantisation,
we have found\cite{CL95c} that it is more convenient to start off on the basis
of a very different kind of variational procedure. 

The standard procedure, as used in Section \ref{Sec2}, for the treatment of
any current, $n^\sigma$ say,  is based on a projection of the four dimensional
space-time background onto a three-dimensional material base space, whose
local coordinates can then be pulled to specify corresponding comoving space
coordinates on the space-time background: the dual three-form,
$\N_{\nu\rho\sigma}$, is obtained as the pullback of a prescribed volume
measure on the base space. In the alternative variation procedure, which has
been shown to lead ultimately to the same equations of motion, both for a
simple perfect fluid\cite{C94a} and for a perfectly conducting two-constituent
model\cite{C94b}, an analogous projection construction is used to specify not
the current three-form $\N_{\nu\rho\sigma}$ but the corresponding vorticity
two form $w_{\rho\sigma}$, which is to be obtained as the pullback of a
prescribed area measure on a two-dimensional base space. In terms of suitably
chosen local coordinates $\chi^{_1}$, $\chi^{_2}$ on the vorticity base manifold, 
the corresponding pair of scalar fields induced on the four dimensional 
spacetime background will specify the vorticity according to the formula
$w_{\rho\sigma}=2\chi^{_1}_{\, ,[\rho}\chi^{_2}_{\, ,\sigma]}$. Such a
prescription automatically ensures that the vorticity two-form will satisfy
both the algebraic degeneracy condition
 \beq
w_{[\mu\nu}w_{\rho\sigma]}=0\, ,
 \eqn{60}\eeq
and the closure condition
 \beq
\nabl_{[\nu}\,w_{\rho\sigma]}=0\,.
 \eqn{61}\eeq

For the purposes of the variational principle it is to be postulated that the
projection of the four dimensional space-time background onto the two
dimensional vorticity base space should be freely variable, which is
equivalent to the postulate that $\chi^{_1}$ and $\chi^{_2}$ should be
considered as freely variable local fields on spacetime. As for the three
dimensional projection used for the direct specification of a current, so
also for the two dimensional projection used for the specification of a
vorticity form, it is convenient to represent an infinitesimal variation via
an infinitesimal displacement vector field, $\xi^\rho$ say, in terms of which
the corresponding base coordinate variations will be given by
$\delta\chi^{_1}=\xi^\rho\chi^{_1}_{\, ,\rho}$,
$\delta\chi^{_2}=\xi^\rho\chi^{_2}_{\, ,\rho}$. As in the case of currents,
the use of such displacement vector fields to specify the corresponding
``convective'' variations enables us to employ a formalism that is manifestly
invariant with respect to changes of the base coordinates, whose explicit use can
thereby be avoided  in the work that follows. It can easily be seen that a
base coordinate independent expression for the infinitesimal variation induced
by the displacement field $\xi^\rho$ will be given by the formula
 \beq
\delta w_{\mu\nu}=-2\nabl_{[\mu}\big( w_{\nu]\rho} \xi^\rho\big)\, .
 \eqn{62}\eeq
This is evidently interpretable, taking account of the closure property
(\ref{60}), as the condition that the infinitesimal variation of $w_{\mu\nu}$
should be given simply by its Lie derivative with respect to the $\xi^\mu$.

The foregoing procedure makes it possible to treat vorticity as an independent
field in its own right. However in such an approach the essential identity of
the vorticity two form as the exterior derivative of a corresponding momentum
one-form must be ultimately be established as a dynamical equation, instead of
being imposed in advance as a defining property in the manner used in the
preceding sections. In order to obtain the required exterior differential
relation between momentum and vorticity in a variational treatment, it has
been found convenient\cite{C94a,C94b,CL95c} to replace the standard
three-dimensional base projection procedure used (as described above) for the
specification of the relevant current, $n^\rho$ say, by an alternative
Kalb-Ramond type procedure whereby the current is specified in terms of a
freely variable gauge two-form, i.e.  an antisymmetric tensor field
$B_{\mu\nu}=-B_{\nu\mu}$ in terms of which the relevant current three-form is
given by
 \beq
\N_{\mu\nu\rho}= 3\nabl_{[\mu}\, B_{\nu\rho]}\, .
 \eqn{64}\eeq
By the Poincar\'e lemma and its inverse, the existence of such a gauge
potential is a sufficient and locally necessary condition for the
closure property
 \beq 
\nabl_{[\mu}\,\N_{\nu\rho\sigma]}=0\, ,
 \eqn{65}\eeq
which is equivalent to the conservation law
 \beq
\nabl_\rho\, n^\rho=0\, .
 \eqn{66}\eeq
The equivalent statement in dual terminology is that this conservation law
(\ref{66}) is the necessary and sufficient condition for the local existence of a
bivector field $b^{\mu\nu}=-b^{\nu\mu}$ such that
 \beq 
n^\rho=\nabl_\sigma\, b^{\rho\sigma}\, ,
 \eqn{67}\eeq
where $b^{\mu\nu}$ is the dual of a corresponding Kalb-Ramond form in
terms of which it will be given by
 \beq
b^{\mu\nu}={_1\over^2}\varepsilon^{\mu\nu\rho\sigma}B_{\rho\sigma}\, .
 \eqn{68}\eeq

It is to be noted that a Kalb-Ramond representation of the form (\ref{64}),
or equivalently (\ref{67}), can not be used in cases
such that of the globally applicable ``transfusive'' model\cite{LSC97} in
which the relevant currents are not separately conserved. It is also to be
remarked that (assuming it is admissible) such a representation will evidently
not be unique: it is subject to gauge transformations specifiable in terms of
an arbitrary one-form $\alpha_{\nu}$ by
 \beq
B{_{\mu\nu}} \, \mapsto \, B{_{\mu\nu}}+ 
2\nabl_{[\mu}\alpha{_{\nu]}}\, .
 \eqn{69}\eeq

In  the particular case of the three constituent models with which
the present discussion is concerned, the considerations presented
above lead us to introduce an isotopic doublet of free gauge bivector fields
$\bn^{\,\rho\sigma}$ and $\bp^{\,\rho\sigma}$ in terms of which
the neutron and proton currents will be given by the combined expression
 \beq
\nU^{\,\rho}=\nabl_\sigma\, \bU^{\,\rho\sigma}\, , 
 \eqn{70}\eeq
where, as in the previous section, it is to be understood that the isotopic
index runs over the values ${\mit \Upsilon}$=n and ${\mit \Upsilon}$ = p.
However since the problem of vorticity quantisation does not arise for the
normal constituent we shall continue to treat it as before, meaning that 
$\ne^{\,\rho}$ will be supposed to be defined, via a projection, as the 
dual of the pullback of the measure on a three dimensional base space.
Treating the gauge fields $\bn^{\,\rho\sigma}$ and $\bp^{\,\rho\sigma}$
as freely variable, but treating the associated vorticities
$\wn_{\,\rho\sigma}$ and $\wp_{\,\rho\sigma}$, via independent
projections, as pullbacks of the measure on a two dimensional base space
in the manner described above, in a new Lagrangian obtained from
(\ref{9}) by a replacement
 \beq
{\cal L}\ \mapsto {\cal L}_{\rm I\! I}\, ,
 \eqn{71}\eeq
where (summing as usual over the isotopic index ${\mit\Upsilon}$)
 \beq
{\cal L}_{\rm I\! I}={\cal L}+{_1\over^2}\bU^{\,\sigma\rho}
\wU_{\,\rho\sigma}\, ,
 \eqn{72}\eeq
it can be verified that one obtains a variational formulation leading back to
field equations precisely equivalent to those obtained by the original
procedure in Section \ref{Sec2}.

The advantage of the new procedure is that, following the example of our
single-constituent prototype\cite{CL95c}, it can easily be generalised so as
to provide a macroscopic description of the effects (whose analysis in a
Newtonian framework was pioneered by Bekarevich and
Khalatnikov\cite{BekaK61,Sonin87}) to be expected as a consequence of vortex
quantisation, by replacing the original formula (\ref{10}) for the gauge
independent action contribution $\Lambda$ in the new Lagrangian
 \beq
{\cal L}_{\rm I\! I}=\Lambda+j^\rho A_\rho
+{_1\over^2}\bU^{\,\sigma\rho}\wU_{\,\rho\sigma}\, ,
 \eqn{74}\eeq
by a more general formula according to which $\Lambda$ is given as an
algebraic function not just of the (gauge independent) fields $\nn^{\,\rho}$,
$\np^{\,\rho}$, $\ne^{\,\rho}$, $s$, $F_{\mu\nu}$, and of course the metric
$g_{\mu\nu}$, as before, but also on the (similarly gauge independent) doublet
of vorticity two-forms, $\wn_{\,\rho\sigma}$ and $\wp_{\,\rho\sigma}$.  This
means that its most general infinitesimal variation will be given by an
expression of the form
 \beq
\delta\Lambda=\muU_{\,\rho}\,\delta \nU^{\,\rho} +{_1\over^2}
\lamU^{\sigma\rho}\,\delta\wU_{\rho\sigma}
+\mue_{\,\rho}\,\delta\ne^{\,\rho}-\Theta\delta s  
+{1\over 8\pi}\H^{\sigma\rho}\,\delta F_{\rho\sigma}
+{\partial \Lambda\over\partial g_{\rho\sigma}}\,\delta g_{\rho\sigma}\, ,
\eqn{75}\eeq
where, as in (\ref{5}), the coefficients of the metric variations are 
not independent of the others but by the relevant generalisation of the
Noether identity (\ref{6}) must satisfy 
 \beq
{\partial \Lambda\over\partial g_{\rho\sigma}}=
{\partial \Lambda\over\partial g_{\sigma\rho}}=
{_1\over^2}\muX_{\,\nu} \nX^{\,\rho}g^{\nu\sigma} +
{_1\over^2}\lamU^{\,\nu\rho}\wU_{\,\nu}{^\sigma}
+{1\over 16\pi}\H^{\nu\rho} F_\nu{^\sigma}\, .
 \eqn{76}\eeq
The new doublet of bivectorial coefficients $\lamU^{\,\sigma\rho} =
-\lamU^{\,\rho\sigma}$ in this expansion characterises the macroscopic
anisotropy arising respectively from the concentration of energy and tension
in mesoscopic  vortices of the neutron and proton superfluids as a
consequence of their vorticity quantisation conditions, in the manner
discussed in our previous work\cite{CL95c} on the single constituent model.
These new four dimensional bivectorial coefficients replace the three
dimensional (space) vectorial coefficients introduced for a similar purpose in
a more restricted Newtonian framework by Bekarevich and
Khalatnikov\cite{BekaK61}. As before, the covectorial coefficients
$\mun_\rho$, $\mup_\rho$, and $\mue_\rho$ will be interpretable respectively
as the mean momentum covectors per particle of the neutrons, the protons, and
the electrons, and $\Theta$ will be interpretable as the temperature of the
``normal constituent''. Finally the bivectorial coefficient $\H^{\rho\sigma}=
-\H^{\sigma\rho}$ will be interpretable  as an electromagnetic displacement
tensor, in terms of which the total electromagnetic field tensor (\ref{11})
will be given by an expression of the form 
 \beq 
F^{\rho\sigma}= \H^{\rho\sigma} + 4\pi \M^{\rho\sigma}\, , 
 \eqn{77}\eeq 
in which $\M^{\rho\sigma}$ is what can be interpreted as the magnetic
polarisation tensor. In the application that we are considering, the --
typically dominant -- polarisation contribution $4\pi\M^{\rho\sigma}$ is to be
thought of as representing the part of the magnetic field confined in the
vortices, while the -- typically much smaller -- remainder $\H^{\rho\sigma}$
represents the average contribution from the external field in between the
vortices, which can be expected to vanish by the ``Meissner effect'' in
strictly static configurations, but which can be expected to acquire a non
zero value  due to the ``London effect'' in rotating configurations.

The analogue of (\ref{28}), as the macroscopic replacement of (\ref{52}), 
for an infinitesimal variation according to the foregoing
rules -- with the neutron and proton vorticity congruences convected
by displacement vector fields $\xin^{\,\rho}$ and $\xip^{\,\rho}$,
and with the flow lines of the ``normal'' constituent convected
by a displacement vector field $\xie^{\,\rho}$ as before --
will be a ``diamond'' variation (i.e. one that includes allowance for the
effect on the background spacetime measure) expressible in the standard form
 \beq
\diamondsuit {\cal L}_{\rm I\! I} =\big(\nabl_\rho\,\piU_{\,\sigma}-
{_1\over^2}\wU_{\,\rho\sigma}\big)\delta \bU^{\,\rho\sigma}+
\fX_{\,\rho}\, \xiX^{\,\rho}
+\Big(j^\rho-{1\over 4\pi}\nabl_\sigma\, \H^{\rho\sigma}\Big)\,\delta A_\rho
+{_1\over^2} T^{\mu\nu}\,\delta g_{\mu\nu} 
+\nabl_\sigma\, {\cal R}_{\rm I\! I}^{\ \sigma} \, ,
 \eqn{80}\eeq
in which the force densities $\fX_{\,\rho}$ appearing
as coefficients of the three independent displacement
displacement vector fields $\xiX^{\,\rho}$ will be given for the isotopic
doublet values {\srm X}=n  and {\srm} X=p by
 \beq
\fn_{\rho} =\big(\nn^{\,\sigma}+\nabl_\nu\,\lamn^{\,\sigma\nu}\big)
\wn_{\, \sigma\rho}\, , \hskip 1 cm    \fp_{\rho} = \big(\np^{\,\sigma}
+\nabl_\nu\,\lamp^{\,\sigma\nu}\big)\wp_{\, \sigma\rho}\, ,
 \eqn{81}\eeq
while the third force density $\fe_{\,\rho}$ acting on the ``normal''
constituent will still be given by the same formula  (\ref{31})
as before. 

Although it is of no relevance for the application of the variation 
principle, it can be noted for the record that the current appearing
in the final divergence term of (\ref{80}) will be given by 
 \beq
{\cal R}_{\rm I\! I}^{\ \sigma}=\piU_{\,\rho}\,\delta\bU^{\,\rho\sigma}
-\big(\bU^{\,\rho\sigma}+\lamU^{\,\rho\sigma}\big)\wU_{\,\rho\nu}\xiU^\nu 
+{_1\over^2}\piU_{\,\rho}\bU^{\,\rho\sigma}
 g^{\mu\nu}\,\delta g_{\mu\nu}+2\Pi_\nu
\xie^{\,[\sigma}u^{\,\nu]}+{1\over 4\pi}H^{\nu\sigma}\,\delta A_\nu\, .       
 \eqn{82}\eeq
An entity of much greater practical interest is the corresponding stress
momentum energy density tensor, which can be seen to be given by
 \beq
T_{\sigma}^{\ \rho}=\nX^{\,\rho}\muX_{\,\sigma}+ 
s\Theta u^\rho u_\sigma+\lamn^{\,\nu\rho}\wn_{\,\nu\sigma}
+\lamp^{\,\nu\rho}\wp_{\,\nu\sigma}+ {1\over 8\pi}\H^{\nu\rho}
F_{\nu\sigma}+\Psi g^\rho_{\ \sigma}\, ,
 \eqn{83}\eeq
where the new generalised pressure function is given by
 \beq
\Psi=\Lambda-\nX^{\,\nu}\muX_{\,\nu}+s\Theta+\bU^{\,\rho\sigma}
\big(\nabl_{\,\rho}\,\piU_{\,\sigma}-{_1\over^2}\wU_{\,\rho\sigma}\big)
 \eqn{84}\eeq
The last term in (\ref{84})  will evidently drop out when we impose the
condition of invariance with respect to infinitesimal variations of the
bivectorial gauge potentials $\bn^{\,\rho\sigma}$ and $\bp^{\,\rho\sigma}$ is
imposed, a requirement which can be seen from (\ref{80}) to give field
equations of the form
 \beq
\wn_{\,\rho\sigma}=2\nabl_{[\rho}\pin{_{\!\sigma]}}
=2\nabl_{[\rho}\mun{_{\!\sigma]}}\, ,\hskip 1 cm\, 
\wp_{\,\rho\sigma}=2\nabl_{[\rho}\pip{_{\!\sigma]}}
=2\nabl_{[\rho}\mup{_{\!\sigma]}}+e F_{\rho\sigma}\, .
 \eqn{85} \eeq
These expressions are evidently equivalent to what in the previous formulation
were merely the definitions of the vorticities as given by the
general specification (\ref{15}).

The remaining field equations obtained from (\ref{80}) will consist of
an electromagnetic source equation having the form
 \beq
\nabl_\sigma\, \H^{\rho\sigma}={4\pi}\, j^\rho\, ,
 \eqn{86} \eeq
together with the condition that the force density coefficients should all
vanish. For the normal constituent, the corresponding condition
 \beq
\fe_{\,\rho}=0\, 
 \eqn{87}\eeq
will take the same form (\ref{44}) as before, but it can be seen
from (\ref{81}) that the other two force balance conditions,
 \beq
\fU_{\,\rho}=0\, ,
 \eqn{88}\eeq
will involve new terms. In particular, the neutron force balance condition
can be seen from (\ref{81}) to be expressible as 
\beq
 2\nn^{\,\sigma}\nabl_{[\sigma}\,\mun_{\rho]}+
\wn_{\,\sigma\rho}\nabl_\nu\,\lamn^{\,\sigma\nu}=0 \, ,
 \eqn{89}\eeq
in which the first term is interpretable as the negative of the Joukowski
force density due to the ``Magnus effect'' acting on the neutron vortices,
while the last term (which was absent in the mesoscopic description)
represents the extra force density on the fluid due to the effect of the
tension of the vortices. The proton force balance equation can be similarly be
seen to be expressible in the form
\beq
2\np^{\,\sigma}\nabl_{[\sigma}\,\mup_{\rho]}+e\np^{\,\sigma} F_{\sigma\rho}
+\wp_{\,\sigma\rho}\nabl_\nu\,\lamp^{\,\sigma\nu}=0  \, ,
 \eqn{90}\eeq
in which the first term is the negative of the Joukowski force density that
acts on the proton vortices, the middle term is the Lorentz force density
representing the effect of the magnetic field on the passing protons, and the
final term again represents the force density due to the effect of the tension
of the vortices.

\section{Entrainment in partially separated models.}
\label{Sec5}

In order to be utilisable for particular applications, the characterisation of
the mesoscopic and macroscopic models described in the preceding sections
needs to be completed by the designation of an appropriate ``equation of
state'' giving the specific form of the mesoscopic $\Lambda_{\rm M}$ and the
macroscopic $\Lambda$ as algebraic functions of the relevant currents and
also, for the latter, of the relevant vorticities as well as the
electromagnetic field. Accurately realistic ``equations of state'' can only be
obtained by detailed microscopic analysis, and even if available would be
likely to be too complicated to be convenient in practice. Since use of a
conservative model involving only three independent constituents is already a
simplification, it is reasonable to employ a correspondingly simplified
``equation of state''.

A very convenient kind of simplification, that is implicitly or
explicitly employed in most of the preceding work in a Newtonian
framework\cite{MenLin91,Men91}, and that will be adopted for the work
of the present section, is to suppose that the action contribution (and
hence the corresponding stress momentum energy contribution) of the
``superfluid'' or ``quantised'' part is separate from that of the
``normal'' or strictly ``classical'' part, meaning that at the level of
the ``mesoscopic'' description the relevant ``matter'' contribution
$\Lambda_{\rm M}$ is taken to be decomposable as a sum in the form
 \beq 
\Lambda_{\rm M}=\Lambda_{\rm Q}+\Lambda_{\rm C}\, ,
 \eqn{101}\eeq 
where, in the three constituent models with which we are concerned here, the
``quantum'' part depends only on the neutron and the proton currents, while
the ``classical'' part of the Lagrangian density depends only on the comoving
electron and entropy currents, so that the variation (\ref{7}) decomposes as
the sum of two separate contributions
 \beq \delta\Lambda_{\rm Q}=\muU_{\ \rho}\, \delta\nU^{\,\rho}+ {_1\over ^2}
\muU_{\ \nu}\, \nU^{\,\rho}\, g^{\nu\sigma}\,\delta g_{\rho\sigma}\, ,
 \eqn{102}\eeq and
 \beq \delta\Lambda_{\rm C}=\mue_{\ \rho}\, \delta\ne^{\,\rho}-\Theta\,
\delta s +{_1\over ^2}\mue_{\ \nu}\, \ne^{\,\rho}\, g^{\nu\sigma}\, 
\delta g_{\rho\sigma}\, .
 \eqn{103}\eeq 
It seems plausible that such a simplification should be fairly
satisfactory for the low temperature limit, $\Theta\rightarrow 0$, thus
providing an adequate approximation for many purposes in neutron star
theory.  However as was clearly understood by Landau when he obtained
the equation of state for the ``cool'' limit of his original two
constituent model\cite{LL59b} (of which an elegant relativistic
analogue has recently been provided by the present authors\cite{CL95a})
there will inevitably be significant coupling between the superfluid
and ``normal'' constituents at higher temperatures, for which the
decomposition (\ref{101}) would therefore cease to be valid.

The simplest conceivable models would be based on not just partial but
complete separation, which would mean that the contribution $\Lambda_{\rm Q}$
in (\ref{101}) would itself split up as the sum of two distinct terms,
$\Lambda_{\rm n}$ and $\Lambda_{\rm p}$ say, of which the first depends only
on the magnitude of the neutron current vector, and the second only on that of
the proton current vector. However ever since the earliest pioneering work by
Andreev and Bashkin\cite{AndreevBashkin76} it has been generally recognised
 that (even in the zero temperature limit) such an extreme simplification
will be unrealistic. This means that the form of $\Lambda_{\rm Q}$ will be
such that the proton and neutron momenta will be given in terms of the
corresponding currents by a (necessarily symmetric) matrix relation of the
form
 \beq
\muU_{\,\rho}= g_{\rho\sigma}\KuUV\nV^{\,\sigma}\, ,
 \eqn{106}\eeq
in which, as well as the diagonal terms $\Kunn$ and $\Kupp$ that would be
present even in the oversimplified fully separated case, there will
also be a cross term $\Kunp=\Kunp$. The presence of such a term implies that
the respective directions of the neutron and proton momenta will deviate from
that of the corresponding currents, in a manner that is familiar in the well
known example of the Landau type two constituent model at non-zero
temperature. Such a (non-dissipative) deviation effect is usually referred to
as {\it entrainment}. (It is however to be noted that there is an alternative
but regrettable usage whereby some authors call it ``drag'', which is
misleading, since the standard use of the latter term is for dissipative
forces opposing relative movement.)  The (symmetric) 2 by 2 matrix
relation (\ref{106}) can be rewritten in inverted form as
 \beq
\nU^{\,\rho}= g^{\rho\sigma}\KdUV\muV_{\,\sigma}\, ,
 \eqn{107}\eeq
using the usual convention that (just as $g^{\rho\sigma}$ denotes the
components of the matrix inverse of the metric tensor $g_{\rho\sigma}$) the
matrix with (lowered) components $\KdUV$ is the inverse of the original matrix
with components $\KuUV$. 

For the purpose of comparison with our treatment\cite{CL95c} of the
single constituent case, for which it was found convenient to work
with a ``dilatonic amplitude'', $\Pf$ defined as the square root
of the ratio of the particle number density to the effective mass,
it is of interest to introduce analogous amplitudes $\Pfn$ and
$\Pfp$ for the protons and neutrons respectively in such a way as
to allow (\ref{107}) to be rewritten more explicitly as
 \beq 
\nn{_\sigma}= \Pfn^{\,2}\,\mun_{\,\sigma}+
\Kdnp(\mup_{\,\sigma}-\mun_{\,\sigma})\, , \hskip 1 cm
\np{_\sigma}= \Pfp^{\,2}\,\mup_{\,\sigma}+
\Kdnp(\mun_{\,\sigma}-\mup_{\,\sigma})\, ,
 \eqn{108} \eeq
which evidently requires the specifications
 \beq
\Pfn^{\,2}=\Kdnn+\Kdnp\, ,\hskip 1 cm 
\Pfp^{\,2}=\Kdpp+\Kdnp\, .
 \eqn{109}\eeq

In order to relate the formula to (\ref{106}) to the expressions
traditionally used in the relevant literature as previously developed in a
Newtonian framework\cite{AndreevBashkin76,ALS84,Men91}, it is convenient to
introduce what from a relativistic point of view is a rather artificial
concept, namely that of the ``rest mass'', $m$ say per baryon, which can
indifferently be taken to be the mass of a free neutron, the (more precisely
definable) mass of a free proton, or as a compromise between the two, the mass
of  an ordinary free hydrogen atom in its ground state, or even the standard
atomic unit as conventionally defined in terms of oxygen. The reason why it
does not matter which of these one uses is that this ``rest mass'' is needed
merely as a unit for calibration, and in any case by the standards of accuracy
of the present treatment all these various alternative definitions give are
virtually the same result. However the calibration mass $m$ is chosen, one
can use it to define corresponding conserved ``mass currents'' $\rhn^{\,\nu}$ 
and $\rhp^{\,\nu}$ for the protons and neutrons by the formula
 \beq
\rhU^{\,\nu}=m\nU^{\,\nu}\, .
 \eqn{112}\eeq
Another popular concept in the Newtonian context, despite the
fact that its calibration is subject to a similar degree of arbitrariness, is
that of what is commonly but misleadingly referred to as a superfluid
``velocity''. One can define such so called ``velocities'' $\vsn{^{\,\nu}}$
and $\vsp{^{\,\nu}}$ for the neutrons and the protons in terms of the chosen
calibration mass $m$ (as specified by one or other of the various conventions
mentioned above for defining the ``baryon rest mass'') by setting  
 \beq
\muU_{\nu} = m\vsU_{\,\nu}\, ,
 \eqn{113}\eeq
which means that $\vsn_{\,\nu}$ and $\vsp_{\,\nu}$ can be read out just as
correspondingly rescaled momentum variables. It is to be observed that, no
matter how the constant mass scale factor $m$ is chosen, $\vsn{^{\,\nu}}$ and
$\vsp{^{\,\nu}}$ can not be made to be ``velocity'' 4-vectors in the strict
sense because their magnitudes will in general be variable, so that they can
not be made to be unit vectors, except in such trivial special cases as that
of a spatially uniform stationary state. In terms of these so called 
``velocities'', which are actually just rescaled momentum variables,
(\ref{107}) can be rewritten in the form
 \beq
\rhU^{\,\rho}= g^{\rho\sigma}\rhUV\vsV_{\,\sigma}\, ,
 \eqn{114}\eeq
where the coefficients $\rhUV$ are components of a ``density matrix'' of the
traditional but unsatisfactorily mass scale dependent kind, which will be
given in terms of the more precisely defined matrix components $\KdUV$
introduced above by 
  \beq
\rhUV=m^2\KdUV\, . 
 \eqn{115}\eeq
Writing this out using the definitions
 \beq
\rhn=\big(m\Pfn\big)^2=\rhnn+\rhnp\, ,\hskip 1 cm
\rhp=\big(m\Pfp\big)^2=\rhpp+\rhnp\, ,
 \eqn{118}\eeq
it can can be be seen that when translated into terms of the
pseudo-velocities, the expansions (\ref{108}) will take the
kind of form that is familiar in the Newtonian literature:
 \beq
\rhn{_\sigma}= \rhn\,\vsn_{\,\sigma}+
\rhnp(\vsp_{\,\sigma}-\vsn_{\,\sigma})\, , \hskip 1 cm
\rhp{_\sigma}= \rhp\,\vsp_{\,\sigma}+
\rhnp(\vsn_{\,\sigma}-\vsp_{\,\sigma})\, .
 \eqn{116}\eeq

\section{Polarisation in partially separated models.}
\label{Sec6}

For the purpose of extending the kind of ``mesoscopic'' model considered
in the previous section to a ``macroscopic'' model representing
the effect of averaging over a congruence of vortex lines
in the manner described in Section \ref{Sec4}, it is useful to
start by using the relations obtained in Section \ref{Sec5}
to obtain rough estimates of the total electromagnetic flux,
 \beq
\Phi=\oint A_\sigma\, dx^\sigma\, ,
 \eqn{125}\eeq
that is to be expected through a circuit at sufficiently large distance
round a vortex defect characterised, according to (\ref{56})
by neutron and proton and winding numbers
$\nun$ and $\nup$. It can be seen, by working out (\ref{107}) for the 
particular case of the proton current $\np^{\,\sigma}$ that the
total value of the electromagnetic gauge field will be expressible
in the form
 \beq
A_\sigma={1\over e}\big(\pip_\sigma+\alpn\pin_\sigma\big) +\Delta_\sigma\, ,
 \eqn{126}\eeq
where the factor $\alpn$ is a quantity that would be zero if there were no 
``entrainment'' but that will in general be given by
 \beq
\alpn={\Kdpn\over\Kdpp}={\rhpn\over\rhpp} \, ,
 \eqn{127}\eeq
while the final remainder term will be given by
\beq
 \Delta^\sigma = -{1\over e\Kdpp}\np{\,^\sigma}
= -{m\over e  \rhpp}\rhp^{\,\sigma}\, .
 \eqn{128}\eeq
Assuming that the coefficients $\KdUV$ tend approximately towards
uniform values at large distance, the corresponding flux $\Phi$
will tend approximately towards a limit given in terms of these values
by
 \beq
\Phi \simeq \nuU\PhiU+\PhiD\, ,
 \eqn{129}\eeq
where, by (\ref{56}), the proton vortex flux contribution will be
given simply by
 \beq
\Phip= {\pi\hbar\over  e}\, ,
 \eqn{130}\eeq
which is just the usual flux quantisation unit associated with the
relevant total charge $2e$ of the Cooper type proton pairs,
while the neutron vortex flux contribution would vanish if there were no
``entrainment'' but will in general be given by the formula
\beq
\Phin={\pi\hbar\over  e}\,\alpn\, ,
 \eqn{131}\eeq
where $\alpn$ is the relevant ``entrainment factor''. The formula 
(\ref{131}), with $\alpn$ given by the final expression in (\ref{127}), is
already familiar from the corresponding analysis in a Newtonian
framework\cite{SS80,ALS84}. If the ``Meissner effect'' were fully effective the
residual flux term would tend to zero, but as was originally realised by
London in the context of ordinary metallic superconductivity this can be
expected only in a strictly static background, whereas more generally even in
a stationary case provided the background is rotating, there will be a
residual magnetic field even at large distances from a vortex, and hence a
corresponding ``London'' flux contribution, which will be given in the present
case by 
 \beq
\PhiD = - {1\over e\Kdpp}\oint \np^{\,\rho} g_{\rho\sigma}\, dx^\sigma\, .
 \eqn{132}\eeq

Under reasonably stable conditions one would expect that energy would
be locally minimised by avoidance of the field build up that would
result from persistent deviation of the proton current contribution
$\np^{\,\sigma}$ from the canceling background current contribution
$\ne^{\,\sigma}$ provided by the ``normal'' electrons, i.e. one would
expect to have
 \beq
\np^{\,\sigma}\approx \ne u^\sigma\, ,
 \eqn{133}\eeq
where $u^\mu$ is the unit 4-velocity vector of the ``normal'' background, as
introduced above. Under such circumstances, the ``London'' flux contribution
(\ref{132}) can be expected to be given in terms of the asymptotic values of
the proton number density $\np$ or the corresponding conventionally normalised
proton mass density as defined by
  \beq
\np=\big(-\np^{\,\sigma}\np{_\sigma}\big)^{1/2}\, ,\hskip 1 cm
\rhp=m\np\, ,
 \eqn{134}\eeq
by an estimate of the form
 \beq
\PhiD\approx -{\np\over e\Kdpp} \oint u_\sigma\, dx^\sigma
= -{m\rhp\over e \rhpp}\oint u_\sigma\, dx^\sigma\, .
 \eqn{135}\eeq

However that may be, the ``London'' flux contribution $\PhiD$ through a
circuit round a plane surface element of area ${\cal S}$ can be attributed to
an average orthogonally oriented magnetic field contribution $\langle
H\rangle$ given by
 \beq
\langle H\rangle ={ \PhiD\over{\cal S}} \, .
 \eqn{136}\eeq
In addition to this external London contribution, the corresponding total
average magnetic field, as defined  by
 \beq
\langle B\rangle={\Phi\over{\cal S}}\, ,
 \eqn{137}\eeq
will contain an internal contribution from the part of the flux that
is confined within the neighbourhood of the vortex. According to
(\ref{129}), this internal contribution will be expressible by
 \beq
\langle B\rangle-\langle H\rangle={1\over\pi\hbar}\PhiU\langle \wU\rangle\, ,
 \eqn{138}\eeq
where the effective average neutron and proton vorticities $\langle\wn\rangle$
and $\langle\wp\rangle$ are given in terms of the corresponding quantised
circulation integrals (\ref{56}) by
 \beq
\langle\wU\rangle={1\over {\cal S}}\oint\piU_{\,\rho}\, dx^{\,\rho}\, ,
 \eqn{139}\eeq

The formula (\ref{138}) provides what one needs for going over from the
``mesoscopic'' level of analysis we have been using so far in this section to
a ``macroscopic'' level of analysis in terms of averages over many vortices.
The average  magnitude $\langle\wU\rangle$ is interpretable as the relevant
component (as determined by the choice of orientation of the circuit under
consideration) of  the kind of macroscopic vorticity tensor $\wU_{\mu\nu}$
that was considered in the preceding section. The total average magnitude
$\langle B\rangle$ is analogously interpretable as the corresponding component
of the macroscopic electromagnetic field tensor $F_{\rho\sigma}$, while the
external ``London'' contribution $\langle H \rangle$ will be similarly
interpretable as the corresponding component of the generalised displacement
field tensor $\H_{\rho\sigma}$ that was introduced in the previous section. By
comparison with (\ref{77}), it can thus be seen that the difference
(\ref{138}) will  be interpretable as the corresponding component of
$4\pi\M_{\rho\sigma}$, where $\M_{\rho\sigma}$ is the relevant polarisation
tensor. It follows that this macroscopic polarisation tensor, representing the
effect of the part of the magnetic field that is confined in the neighbourhood
of the vortices, will be given in terms of the macroscopic proton and neutron
vorticity tensors by
 \beq
4\pi^2\hbar\M_{\rho\sigma}=\PhiU\wU_{\rho\sigma}\, .
 \eqn{150}
\eeq
It is important to point out that this result differs from the result 
obtained in the Newtonian limit by Mendell \cite{Men91}, 
who found  the polarisation to be zero. However that conclusion was 
based on the inclusion of an extra magnetic interaction  term but 
on the neglect of a kinetic contribution that, according to a more
recent analysis \cite{cpl98}, will actually cancel it out.

Working  out the polarisation tensor explicitly in terms of the 
elementary proton and neutron
vortex flux units (\ref{130}) and (\ref{131}) it can be seen that the
Dirac-Planck constant $\hbar$ cancels out, leaving a relation of the form
 \beq
4\pi e \M_{\rho\sigma}=\wp_{\,\rho\sigma}+\alpn \wn_{\,\rho\sigma}\, ,
 \eqn{151}\eeq
in which the last term on the right is entirely due to the ``entrainment''
effect, without which the neutron vorticity would not contribute to the
magnetic polarisation.  According to (\ref{77}) the 
displacement field due to the ``London effect''   will be given by
 \beq
\H_{\mu\nu}=F_{\mu\nu}-{1\over e}\big(\wp_{\,\mu\nu}
+\alpn \wn_{\,\mu\nu}\big) \, .
 \eqn{152}\eeq

In terms of the ``normal'' background's acceleration vector $\dot u^\nu$ and
rotation tensor $\omeg_{\mu\nu}$ (whose magnitude $\omeg=
\big({_1\over^2}\omeg_{\mu\nu}\omeg^{\mu\nu})^{1/2}$ is the local angular
velocity) as defined by
 \beq
\nabl_{[\mu}\,u_{\nu]}=2\omeg_{\mu\nu}- u_{[\mu} \dot u_{\nu]}\, ,
\hskip 1 cm \dot u^\mu= u^\nu\nabl_\nu\, u^\mu\, ,
 \eqn{153}\eeq
one would expect that in steady circumstances such that the estimate
(\ref{135}) is valid, the ``London'' field would be close to a value given by
 \beq
\H_{\mu\nu}\approx {\np\over e\Kdpp}\big(u_{[\mu} \dot u_{\nu]}
-2\omeg_{\mu\nu}\big)={m\rhp\over e\rhpp}\big(u_{[\mu} \dot u_{\nu]}
-2\omeg_{\mu\nu}\big)\, ,
 \eqn{154}\eeq
in which it can be seen that the electric part is proportional to
the acceleration while the magnetic part is proportional to the
angular velocity.

The ``London field'' (\ref{152}) will be interpretable (whether or not the
conditions for (\ref{154}) are satisfied) as representing the contribution to
the total macroscopic average electromagnetic field $F_{\mu\nu}$ from the 
magnetic field (if any) ``outside'' the vortices. Such a heuristic
interpretation should not be taken too seriously because there will not be an
absolutely clear cut boundary between the ``inside'' and the ``outside'' of a
vortex. Nevertheless the distinction between the internal contribution
described in terms of polarisation and the external contribution described as
a  ``London'' field will usually be fairly precise in practice, because  the
magnetic field can be expected to suffer an exponential decline with a lengthscale
small compared to the intervortex separation. 

In so far as there is some residual degree of arbitrariness in the distinction
between the internal ``polarisation'' contribution and the external ``London''
contribution, the formula (\ref{152}) can be considered as an exact defining
condition. However this convention will only be consistent with the original
definition of the displacement field via (\ref{75}) in terms of partial
differentiation with respect to the total field $F_{\mu\nu}$ if the Lagrangian
density contribution $\Lambda$ has the right functional form. For mathematical
consistency between (\ref{75}) and (\ref{152}) it is necessary and sufficient
that $\Lambda$ should be decomposable in the form
 \beq
\Lambda=\Lambda_{\rm MV}+\Lambda_{\rm F}\, ,
 \eqn{160}\eeq
where the macroscopic contribution $\Lambda_{\rm MV}$ is required to be
functionally independent of $F_{\mu\nu}$, though it is dependent now on the
vorticities $\wU_{\,\rho\sigma}$ as well as on the currents, while the
remaining contribution is required to have the specific form
 \beq
\Lambda_{\rm F}={1\over 16\pi} \H_{\rho\sigma}\H^{\sigma\rho}\, .
 \eqn{161}\eeq
So long as the vortices are (as is expected to be the case in typical neutron
star application) sufficiently far apart to justify the heuristic interpretation
of $\H_{\mu\nu}$ as the intervortex contribution to the macroscopically
averaged field $F_{\mu\nu}$, it is physically plausible that the action
contribution of this ``London'' field should be given by substituting
$\H_{\rho\sigma}$ for $F_{\mu\nu}$ in the standard Maxwellian formula
(\ref{12}). Since the formula (\ref{161}) is precisely what results from such
a substitution, the physical coherence of the description we have developed
here is reassuringly confirmed.

In the context of the partial separation approximation on which the analysis
of this section is based, there will be no loss of generality in decomposing
the macroscopic Lagrangian density contribution $\Lambda_{\rm MV}$ in the form
 \beq
\Lambda_{\rm MV}=\Lambda_{\rm M}+\Lambda_{\rm V}\, ,
  \eqn{162}\eeq
where  $\Lambda_{\rm M}$ is the function already used in the mesoscopic
analysis, which in the present section is assumed to have the partially
separated form (\ref{101}), while $\Lambda_{\rm V}$ is an extra term which
unlike $\Lambda_{\rm M}$ is algebraically dependent not just on the currents
but also on the vorticity 2-forms.

Before concluding, it is to be remarked that, for the purpose 
of comparison with previous work in Newtonian theory, it may be useful to 
introduce in our macroscopic model the notions of ``velocities" and of 
``density matrix" that were defined in a mesoscopic context in Section 5.
This will mean to replace the 
(in our opinion
more natural) forms (\ref{89}) and (\ref{90}) of the
respective neutron and proton equations
of motion by alternative decompositions
 \beq  
2\rhn{\vsn}^{\,\sigma}\nabl_{[\sigma}\,\vsn_{\nu]}
=\rhnp \big(\vsp{^\sigma}-\vsn{^\sigma}\big)m^{-1}
\wn_{\nu\sigma}+\wn_{\nu\sigma}\nabl_\rho\,
\lamn^{\sigma\rho}\, ,
 \eqn{170}\eeq
and
 \beq 
2\rhp{\vsp}^{\,\sigma}\nabl_{[\sigma}\,\vsp_{\nu]}
=\rhnp \big(\vsn{^\sigma}-\vsp{^\sigma}\big)m^{-1}\wp_{\nu\sigma}
- {e\over m}\big(\rhpp\vsp{^\sigma}+\rhnp\vsn{^\sigma}\big) F_{\sigma\nu}
 +\wp_{\nu\sigma}\nabl_\rho\,
\lamp^{\sigma\rho}\, ,
 \eqn{171}\eeq
in each of which the first term on the right can be interpreted as an
``entrainment force'' density. It is to be observed that -- due to the
antisymmetry of the generalised vorticity tensors-- such entrainment forces
always act orthogonally to the relevant relative difference
$\vsn{^\sigma}-\vsp{^\sigma}$, thereby respecting the strictly conservative
character of the model. (The occasional use in the literature of the term
``drag'' instead of ``entrainment'' can be seen to be misleading, since a
genuine drag force is intrinsically dissipative,  and if present would not act
orthogonally but on the contrary would be aligned with the relevant relative
difference. In order to conform with standard aerodynamical terminology,
the ``entrainment'' force should be described not as ``drag''
but rather as a kind of ``lift''.)

\section{Conclusions}

What remains for future work is the derivation of a suitable formula for the
explicit form of the contribution $\Lambda_{\rm MV}$. Supposing that the
vorticity independent mesoscopic contributions $\Lambda_{\rm Q}$ and
$\Lambda_{\rm C}$ are known, the ``entrainment ratio'' needed for the
specification via (\ref{161}) of $\Lambda_{\rm F}$ will be given explicitly by
(\ref{127}), and one might hope that (as in the example of our preceding
study of the single constituent case\cite{CL95c}) there might be a similarly
simple and explicit formula for the corresponding form of the extra vorticity
dependent term $\Lambda_{\rm V}$. 

If the vortices were of purely ``local'' type, with not just their magnetic
but also the associated circulating currents falling off exponentially in the
exterior, then one would expect that $\Lambda_{\rm V}$ would just be a
homogeneous linear function the magnitudes, $\wn=\big({_1\over^2}
\wn_{\,\mu}{^\nu}\wn_{\,\nu}{^\mu}\big)^{1/2}$ and $\wp=\big({_1\over^2}
\wp_{\,\mu}{^\nu}\wp_{\,\nu}{^\mu}\big)^{1/2}$ of the corresponding vorticity forms.
In the absence of ``entrainment'' the elementary proton vortices would indeed
be of this ``local'' type (like the electron vortices in ordinary ``type II''
metallic superconductors). However even in the single constituent
case\cite{CL95c} the neutron vortices are of ``global'' type, with energy
logarithmically dependent on their separation, and in the present case the
same will be true of the proton vortices, due to the non zero value that is
expected\cite{ALS84} for the entrainment factor $\alpn$ given by (\ref{127}).
The simultaneous presence of neutron and proton vortices can give rise to
interesting interaction effects, which have been examined in a Newtonian
analysis by the Sedrakians\cite{SS95}. However the most suitable way to
represent these effects explicitly within the present framework is not yet
clear to us.

Another important issue would be  to work out  the  magnetohydrodynamic limit of 
our equations of motion in order to get rid of superfluous degrees of freedom 
and to obtain more tractable equations when one wishes to consider 
some specific problems such as glitches in neutron stars.

\section{Acknowledgements}

We would like to thank L. Lindblom and D. Sedrakian for very stimulating 
discussions.


\end{document}